# Leakage radiation microscopy of surface plasmons launched by a nanodiamond-based tip

Oriane MOLLET, Aurélien CUCHE, Aurélien DREZET, Serge HUANT

Institut Néel, CNRS & Université Joseph Fourier, BP 166, 38042 Grenoble cedex, France

* Corresponding author: oriane.mollet@grenoble.cnrs.fr (O. Mollet)

**ABSTRACT**: Leakage-radiation microscopy of a thin gold film demonstrates the ability of an ensemble of fluorescent diamond nanoparticles attached onto the apex of an optical tip to serve as an efficient near-field surface-plasmon polariton launcher. The implementation of the nanodiamond-based tip in a near-field scanning optical microscope will allow for an accurate control on the launching position, thereby opening the way to scanning plasmonics.

*Keywords: nanodiamonds, nitrogen-vacancy (NV) color centers, plasmonics*

**I. Introduction**

The miniaturization of electronic and optical devices faces the crucial obstacle of energy transfer and communication between functional objects at the nanoscale. In the last decade, the use of plasmonics potentiality has been initiated to challenge this issue. Surface-plasmon polaritons (SPPs) [1, 2], i.e. hybrid electron-photon modes propagating at the interface between a metal and a dielectric material, are naturally well adapted to low-dimensional operation. Indeed, as surface waves, SPPs are exponentially damped in the perpendicular direction and can be guided over microns in structures laterally miniaturized to dimensions falling well below the diffraction limit [1, 2]. These appealing properties make SPPs promising candidates for nanoscale optical addressing [3] or light manipulation by tailored plasmonic components, such as beam splitters or multiplexers [4, 5].

Previous studies have shown that SPPs can be launched in a metal film by dipping into its near field an optical tip for near-field scanning optical microscopy (NSOM) coupled to an excitation source of appropriate wavelength (e.g. red light for gold) [6-8]. This is because the near-field (evanescent) light diffracted at the sub-wavelength tip apex possesses high spatial frequencies able to couple to SPPs. This is in contrast with far-field (propagating) photons, which do not have enough momentum. Here we show that the near-field fluorescence emanating from diamond nanocrystals attached onto an optical tip can also efficiently launch SPPs on a metal film. This should allow for a tighter spatial control on the SPP injection position and opens the way to deterministic plasmonics, or even quantum plasmonics if a single diamond nanocrystal hosting a single or a few quantum emitters is used as SPP launcher.

**II. Experimental details**

## II.1. Nanodiamonds

Thanks to their unique long-term photostability and cytocompatibility, fluorescent diamond particles of size below 50 nm [9-12], hereafter referred to as *nanodiamonds* (NDs), have shown a great potential in many applications such as biological markers and trackers [13-15], quantum optics and nano-optics, where they can be used as room temperature single-photon emitters [16] or non-blinking and non-bleaching nano-emitters [17-20], and high-sensitivity high-resolution magnetic sensors [21,22].

The NDs used in this work are the same as in [19,20]. They are obtained from type Ib synthetic diamond powder with particle size centered around 25 nm and Nitrogen content of 100 ppm. This powder is proton-irradiated (2.4 MeV, dose $5.10^{16}$ $H^+/cm^2$) to generate vacancies in the diamond lattice. Thermal annealing activates the migration of vacancies towards nitrogen impurities, thereby forming *Nitrogen-Vacancy (NV) color-centers*. Acid treatment and washing with water lead to a stable aqueous suspension. Intense ultra-sonification disperses the particles at their primary size and results in a very stable colloidal suspension. Due to the acid treatment, NDs bear negatively charged carboxylic groups on their surface. Photon-intensity time correlation measurements [18] reveal that most NDs used here host two NV centers, or more, with a small fraction hosting one color-center only [19].

## II.2. Nanodiamond-based active tips as SPP launchers

The first step of the experiment consists in attaching NDs onto the apex of an optical tip. For this purpose, we follow the method that we have described in detail in [20]. Briefly, an optical tip covered with a thin layer of a positively charged polymer (poly-l-lysine) is raster scanned at short distance (approximately 20 nm) above a fused silica cover slip where negatively charged NDs have been spin cast. Thanks to the electrostatic interaction between the negatively charged NDs and the positively charged polymer, NDs, in the number of five in the present experiment, are embarked during scanning as confirmed by the NSOM fluorescence image of Fig. 1(a) and the corresponding intensity crosscut of Fig. 1(b).

The "graft" is strong enough with poly-l-lysine and the NDs are simultaneously exceptionally photostable so that the functionalized tip can be used for subsequent optical experiments on a long term (typically several days before loosing the NDs), in contrast with ND attachments achieved onto uncoated optical tips [23] or with tips stained with blinking semiconductor quantum dots [24]. Optical characterization of the functionalized tip can be carried out such as photon-intensity time correlation measurements [19] (this is however not relevant in the present work which is dealing with a rather large number of NDs hosting in addition 2 NVs for most of them) or spectroscopic measurements; see Fig. 1(c). The spectrum of Fig. 1(c) is typical for neutral $NV^0$ centers [25], which are usually favored in crystals of size around 20 nm [9,26]. Converting these color-centers to their negatively charged counterparts $NV^-$ in ultrasmall (size below 30 nm) nanodiamonds requires a special treatment as reported in [27].

In the rest of this paper, we use the five-ND functionalized tip of Fig. 1 to launch SPPs on a thin (60 nm) gold film deposited on a fused silica cover slip. We expect this launching to be efficient in the near field because the NV emission falls in the red-orange part of the spectrum, which is appropriate for exciting SPPs in gold, and because the fluorescence light has large spatial frequencies in the near-field of the NDs (next section).

## II.3. Leakage Radiation Microscopy

SPPs propagating at the air-gold interface are characterized by a complex in-plane wavevector

$$k_{SPP}(\lambda) = \frac{2\pi}{\lambda} \sqrt{\frac{\varepsilon_G(\lambda)}{\varepsilon_G(\lambda)+1}} = \frac{2\pi}{\lambda} n_{SPP}(\lambda), \qquad (1)$$

which depends on the optical wavelength $\lambda$ and on the dielectric permittivity $\varepsilon_G(\lambda)$ of gold (which also depends on $\lambda$). This dispersion relation is equivalently defined by the complex SPP optical index $n_{SPP}(\lambda)$. As it turns out, $\mathrm{Re}[k_{SPP}(\lambda)] > \frac{2\pi}{\lambda}$ (Re: real part), therefore preventing far-field photons with total momentum $\frac{2\pi}{\lambda}$ to couple to SPPs. One way of circumventing this limitation is to add a periodic grating on the metal film [28,29]. Another way is to go to the near field to take advantage of the high spatial frequencies available there [6]. This is what we do in the present work.

Due to boundary conditions and conservation of the in-plane wavevector along the different interfaces, SPPs leak through the thin gold film (Fig. 2) into the glass substrate (with optical index $n_{glass}$) at a leakage-radiation (LR) angle given by:

$$\sin[\Theta_{LR}(\lambda)] \cong \mathrm{Re}[n_{SPP}(\lambda)] / n_{glass} \qquad (2).$$

Leakage-radiation microscopy (LRM) [6,30,31] consists in detecting these leaky waves and imaging them directly on a CCD camera. LRM works if the metal film thickness is not much larger than the SPP penetration depth in the metal (i.e.

$\delta z \approx 10$ nm in the visible), which is still the case for a 60 nm thick gold film illuminated with the red-orange light emitted by NV centers [25]; see Fig. 1c.

A scheme of our LRM set-up, directly inspired from [6,30,31] where more details can be found, is shown in Fig. 3. It consists in a set of lenses and filters used to image different variants of the gold film NSOM image available in the (NSOM) image plane. For instance, implementing the removable lens $L_2$ so that its object focal plane coincides with the back focal plane of $L_1$ - which defines the NSOM Fourier plane - provides SPP images that are Fourier plane (FP) images, whereas removing $L_2$ simply provides "Image Plane" (IP) images of the SPP propagation. In addition, implementing a beam-blocker of appropriate extension (see next section) in the Fourier plane allows for filtering the short spatial-frequency components of the transmitted light - the so-called allowed light - without affecting the high-frequency components, i.e. the so-called forbidden light to which SPPs belong (because $\text{Re}[k_{SPP}(\lambda)] > \frac{2\pi}{\lambda}$, see also next section). Hence, with a single set-up, it is possible to obtain IP and FP images of the SPP propagation, both kinds of images being either Fourier filtered (labeled F in this paper) or unfiltered (labeled U), depending on whether the Fourier filter is set or not in the detection path, respectively. We will see below that this remarkable completeness makes of LRM a very powerful tool for characterizing the SPP propagation.

**III. Results and discussion**

Fig. 4 displays a set of LRM images of a 60 nm thick homogeneous gold film taken in the NV emission band (a band-pass filter transmitting light in the $\lambda \in [572 \ nm, 642 \ nm]$ is included in the optical path) that proves that the ND-based active tip of Fig. 1 is able to launch SPPs on the metal film when it is brought in its near field. Note that the excitation light at 488 nm wavelength used for turning on the NV-center fluorescence is unable to excite SPPs in gold due to strong inter-band absorption [32].

The upper panel in Fig. 4, respectively lower panel, stands for images in the real space (IP images), respectively in the Fourier space (FP images). The latter are all unfiltered in the Fourier space (U images) whereas the IP images are all Fourier filtered (F images) except for the insert of Fig. 4(c), which is unfiltered. Figs. 4(b,e) are taken with the tip placed in the near field of the gold film and should be compared on one hand with Figs 4(a,d), which are recorded with the tip kept in the far-field above the gold film, and, on the other hand, with Figs. 4(c,f), which are obtained with the tip placed in the near-field of a metal-free (SPP-free) zone of the glass cover slip.

A salient feature of the near-field IP image in Fig. 4(b) is the signal "explosion" which extends tens of microns away from the tip position and translates into a well-defined circle in the corresponding FP image in Fig. 4(e). This behavior is typical for the SPP propagation [6,30,31] taking place in the direct space over a (propagation) length [6,33] $L_{SPP} = 1/(2\,\text{Im}[k_{SPP}]) \approx 10-20 \mu m$ (Im: Imaginary part) and translating in the Fourier space into a circle of radius $\delta k = \text{Re}[k_{SPP}]$. Clearly, this behavior is neither observed in the far-field of gold (Figs. 4(a,d)) nor in the near-field of glass (Figs. 4(c,f)).

The images in Fig. 4 give us the opportunity of explaining how the Fourier filter is defined in our experiment. When the tip is maintained in the far field of the glass substrate (not shown), the only detected light is the ND fluorescence allowed light transmitted through the substrate at angles smaller than $\Theta_c = \arcsin(1/n_{glass}) \approx 42°$, with $n_{glass} \approx 1.5$. This allowed light produces a circle in the FP that defines the lateral extension of the Fourier filter as drawn by the dash-dotted circles in Figs. 4(e,f). The effect of this Fourier filter is clearly shown in Fig. 4(c) where the straight spot of the U image in the insert has disappeared in the main F image. It is worth stressing that the Fourier filter has (fortunately) no effect on the SPP circle in Fig. 4(e). This is because $\Theta_{LR} > \Theta_C$ as seen from Eq. (2) where $n_{glass} > \text{Re}[n_{SPP}] > 1$. Note however that in a near-field image recorded with the tip facing glass, Fig 4(f), some light is observed outside the Fourier circle. It is due to evanescent waves emanating from the tip with spatial frequencies larger than $2\pi/\lambda$ that couple into the substrate with refraction angles larger than $\Theta_C$, even larger than $\Theta_{LR}$ actually.

A distinctive feature of the SPP circle in Fig. 4(e) is its narrowness. This may seem surprising at first sight since the SPP wavevector is wavelength dependent (Eq. (1)) and the NV emission is rather broad (Fig. 1(c)). The explanation here is that the wavelength dependence of $k_{SPP}$, i.e. of $\varepsilon_G$, in Eq. (1) partly counterbalances the direct wavelength dependence of $\Theta_{LR}$ in Eq. (2) in the relevant wavelength range of the NV emission, which is additionally further reduced by the use of a band-pass filter. This is confirmed in Fig. 5 where we plot the leakage-radiation angle as function of wavelength. Here, the dielectric function of gold has been taken from [32]. As can be seen from Fig. 5, $\Theta_{LR}$ falls within the narrow {44°, 45.5°} angular interval in the relevant wavelength range, which results in the narrow FP circle in Fig. 4(e).

Additional more technical points in Fig. 4 are worth a short comment. First, the allowed-light in Fig. 4(d) is slightly distorted. This is a consequence of an imperfect positioning of the Fourier plane in this particular image. Second, the IP image of Fig. 4(b) is perturbed by an interference fringe pattern resulting in a dark central spot where the highest SPP intensity is expected [6]. This is an artifact, which we believe is due to a small optical index mismatch between the coverslip and the immersion oil.

**IV Conclusions and perspectives**

We have shown by means of leakage radiation microscopy that the near-field fluorescence of NV centers emanating from a five nanodiamond-based optical tip is able to launch SPPs on a gold film. Because the tip is implemented in a NSOM setup, the SPP launcher can be positioned in all three dimensions with nanometer accuracy. This opens the way to a "deterministic" scanning plasmonics. Reducing the number of active nanodiamonds down to one with a few (one or two) NV-center occupancy [34] will not only improve the accuracy in the launching position but will in addition open the way to a "deterministic" quantum plasmonics where the scanning quantum source of light launches one single SPP at a time. This should open the way to various studies in quantum plasmonics with unprecedented accuracy such as coupling of single emitters to single SPPs [35,36] or wave-particle duality for SPPs [37].

**ACKNOWLEGMENTS**

This work is partly supported by the Agence Nationale de la Recherche through the NAPHO and PlasTips projects. We are grateful to J.-F. Roch, F. Treussart, V. Jacques and O. Arcizet for encouraging discussions on this work, to J.-P. Boudou and T. Sauvage for the nanodiamond sample and to J.-F. Motte and T. Fournier (NANOFAB facility) for the optical tip and gold film preparation.

**References**

[1] W. L. Barnes, A. Dereux, T. W. Ebbesen, Nature 424 (2003) 824.

[2] L. Novotny, B. Hecht, Principles of Nano-Optics, Cambridge Press, London, 2006.

[3] A. Drezet, A. Hohenau, J. R. Krenn, M. Brun, S. Huant, Micron, 38 (2007) 427.

[4] A. Hohenau, J. R. Krenn, A.L. Stepanov, A. Drezet, H. Ditlbacher, B. Steinberger, A. Leitner, F. R. Aussenegg, Opt. Lett. 30 (2005) 893.

[5] A. Drezet, D. Koller, A. Hohenau, A. Leitner A, F. R. Aussenegg, J. R. Krenn, Nano Lett. 7 (2007) 1697.

[6] B. Hecht, H. Bielefeldt, L. Novotny, Y. Inouye, D. W. Pohl, Phys. Rev. Lett. 77 (1996) 1889.

[7] C. Sönnichsen, A. C. Duch, G. Steiniger, M. Koch, G. von Plessen, J. Feldmann, Appl. Phys. Lett. 76 (2000) 140.

[8] M. Brun, A. Drezet, H. Mariette, N. Chevalier, J. C. Woehl, S. Huant, Europhys. Lett. 64 (2003) 634.

[9] F. Treussart, V. Jacques, E. Wu, T. Gacoin, P. Grangier, J.-F. Roch, Physica B 376-377 (2006) 926.

[10] J.-P. Boudou, P. A. Curmi, F. Jelezko, J. Wrachtrup, P. Aubert, M. Sennour, G. Balasubramanian, R. Reuter, A. Thorel, E. Gaffet, Nanotechnology 20 (2009) 235602.

[11] G. Dantelle, A. Slablab, L. Rondin, F. Laine, F. Carrel F, P. Bergonzo, S. Perruchas, T. Gacoin, F. Treussart, J.-F. Roch, J. Lumin. 130 (2010) 1655.

[12] O. Faklaris, J. Botsoa, T. Sauvage, J.-F. Roch, F. Treussart, Diamond Relat. Mater. 19 (2010) 988.

[13] C.-C. Fu, H.-Y. Lee, K. Chen, T.-S. Lim, H.-Y. Wu, P.-K. Lin, P.-K. Wei, P.-H. Tsao, H.-C. Chang, W. Fann, Proc. Natl. Acad. Sci. USA 104 (2007) 727.

[14] Y.-R. Chang, H.-Y. Lee, K. Chen, C.-C. Chang, D.-S. Tsai, C.-C. Fu, T.-S. Lim, Y.-K. Tzeng, C.-Y. Fang, C.-C. Han, H.-C. Chang, W. Fann, Nature Nanotech. 3 (2008) 284.

[15] O. Faklaris, D. Garrot, V. Joshi, F. Druon, J.-P. Boudou, T. Sauvage, P. Georges, P.A. Curmi, F. Treussart, Small 4 (2008) 2236.

[16] A. Beveratos, S. Kühn, R. Brouri, T. Gacoin, J.-P. Poizat, P. Grangier, Eur. Phys. J. D 18 (2002) 191.

[17] J. R. Rabeau, A. Stacey, A. Rabeau, F. Jelezko, I. Mirza, J. Wrachtrup, S. Prawer, Nano Lett. 7 (2007) 3433.

[18] Y. Sonnefraud, A. Cuche, O. Faklaris, J.-P. Boudou, T. Sauvage, J.-F. Roch, F. Treussart, S. Huant, Opt. Lett. 33 (2008) 611.

[19] A. Cuche, A. Drezet, Y. Sonnefraud, O. Faklaris, F. Treussart, J.-F. Roch, S. Huant, Opt. Express 17 (2009) 19969.

[20] A. Cuche, A. Drezet, J.-F. Roch, F. Treussart, S. Huant, J. Nanophoton. 4 (2010) 043506.

[21] C. L. Degen, Appl. Phys. Lett. 92 (2008) 243111.

[22] G. Balasubramanian, I. Y. Chan, R. Koselov, M. Al-Hmoud, J. Tisler, C. Shin, C. Kim, A. Wojcik, P. R. Hemmer, A. Krueger, T. Hanke, A. Leitenstorfer, R. Bratschitsch, F. Felezko, J. Wrachtrup, Nature 455 (2008) 648.

[23] A. Cuche, Y. Sonnefraud, O. Faklaris, D. Garrot, J.-P. Boudou, T. Sauvage, J.-F. Roch, F. Treussart, S. Huant, J. Lumin. 129 (2009) 1475.

[24] N. Chevalier, M. J. Nasse, J. C. Woehl, P. Reiss, J. Bleuse, F. Chandezon, S. Huant, Nanotechnology 16 (2005) 613.

[25] Y. Dumeige, F. Treussart, R. Alléaume, T. Gacoin, J.-F. Roch, P. Grangier, J. Lumin. 109 (2004) 61.

[26] C. Santori, P. E. Barclay, K. M. C. Fu, R. G. Beausoleil, Phys. Rev. B 79 (2009) 125313.

[27] L. Rondin, G. Dantelle, A. Slablab, F. Treussart, P. Bergonzo, S. Perruchas, T. Gacoin, H.-C. Chang, V. Jacques, J.-F. Roch, Phys. Rev. B 82 (2010) 115449.

[28] U. Schröter, D. Heitmann, Phys. Rev. B 58 (1998) 15419.

[29] C. Genet, T. W. Ebbesen, Nature 445 (2007) 39.

[30] A. Drezet, A. Hohenau, D. Koller, A. Stepanov, H. Ditlbacher, B. Steinberger, F. R. Aussenegg, A. Leitner, J. R. Krenn, Mater. Sci. Eng. B 149 (2008) 220.

[31] A. Drezet, A. Hohenau, A. L. Stepanov, H. Ditlbacher, B. Steinberger, N. Galler, F. R. Aussenegg, A. Leitner, J. R. Krenn, Appl. Phys. Lett. 89 (2006) 091117.

[32] P. B. Johnson, R. W. Christy, Phys. Rev. B 6 (1972) 4370.

[33] This estimate of the SPP propagation length $L_{SPP}$ is deduced from the intensity crosscut of Fig. 3c, with the SPP intensity scaling like $\frac{\exp(-r/L_{SPP})}{r}$ at distance r from the launching position.

[34] A. Cuche, O. Mollet, A. Drezet, S. Huant, Nano Lett. 10 (2010) 4566.

[35] A. V. Akimov, A. Mukherjee, C. L. Yu, D. E. Chang, A. S. Zibrov, P. R. Hemmer, H. Park, M. D. Lukin, Nature 450 (2007) 402.

[36] S. Gerber, F. Reil, U. Hohenester, T. Schlagenhaufen, J.R. Krenn, A. Leitner, Phys. Rev. B 75 (2007) 073404.

[37] R. Kolesov, B. Grotz, G. Balasubramanian, R. J. Stöhr, A. A. L. Nicolet, P. R. Hemmer, F. Jelezko, J. Wrachtrup, Nature Phys. 5 (2009) 470.

**Figure captions:**

**Figure 1:** (a) Fluorescence image acquired in the spectral range of the NV emission showing the successive trapping of five fluorescent NDs during the tip scan at short distance (≈ 20 nm) above the substrate (a fused silica cover slip). Short arrows tentatively point towards the trapping locations. Each trapping event translates into a persistent increase in the tip-ND assembly fluorescence. Un-trapped diamonds produce distinctive fluorescence spots on the image. Experimental parameters are: 5 x 5 µm$^2$ scanning area, excitation at 488 nm, 100 µW excitation power at the tip apex, 50 ms acquisition time per pixel. (b) Intensity crosscut along the dash-dotted arrow of a) (parallel to the slow-scan direction) showing five successive trapping events labeled +1. (c) Fluorescence spectrum (collection time: 180 s, excitation power: 100 µW) of the functionalized tip confirming the presence of neutral NV$^0$ centers at the tip apex with a distinctive zero-phonon line at 575 nm.

**Figure 2:** Leakage of SPPs launched by a point-like source (S) into the glass substrate. $\Theta_{LR}$ and $\Theta_C$ are the leakage radiation angle and the critical incidence for total internal reflection in glass, respectively. Note that $\Theta_{LR} > \Theta_C$ as explained in the text.

**Figure 3:** Scheme of the LRM set-up used in this work. NSOM stands for the optical tip with NDs at the apex; SPP: surface-plasmon polaritons; MO : oil-immersion microscope objective (NA=1.4, X100). L designates a set of lenses and other optical components that are internal to the inverted microscope on which we have built the NSOM head. There is no access to these components. $L_1$, $L_2$, and $L_3$ are achromatic doublets. The images are directly obtained on a CCD camera.

**Figure 4:** LRM images of a 60 nm thick gold film taken on a CCD camera in the $\lambda \in [572\ nm, 642\ nm]$ wavelength range with the five-nanodiamond based optical tip of Fig. 1 used for illumination. (a-c) are direct space IP (Image Plane, see Fig. 3) images; (d-f) are Fourier space FP (Fourier Plane) images. F (respectively U) stands for Fourier filtered

(respectively unfiltered) images depending on the presence (respectively removal) of the Fourier filter of Fig. 3. (a) and (d) are taken with the tip in the far-field of gold. (b) and (e) are taken with the tip in the near-field of gold. (c) and (f) are taken with the near-field of glass.

**Figure 5:** Leakage radiation angle computed as function of wavelength for the air-gold interface. The dashed zone marks the band-pass filter used for imaging.

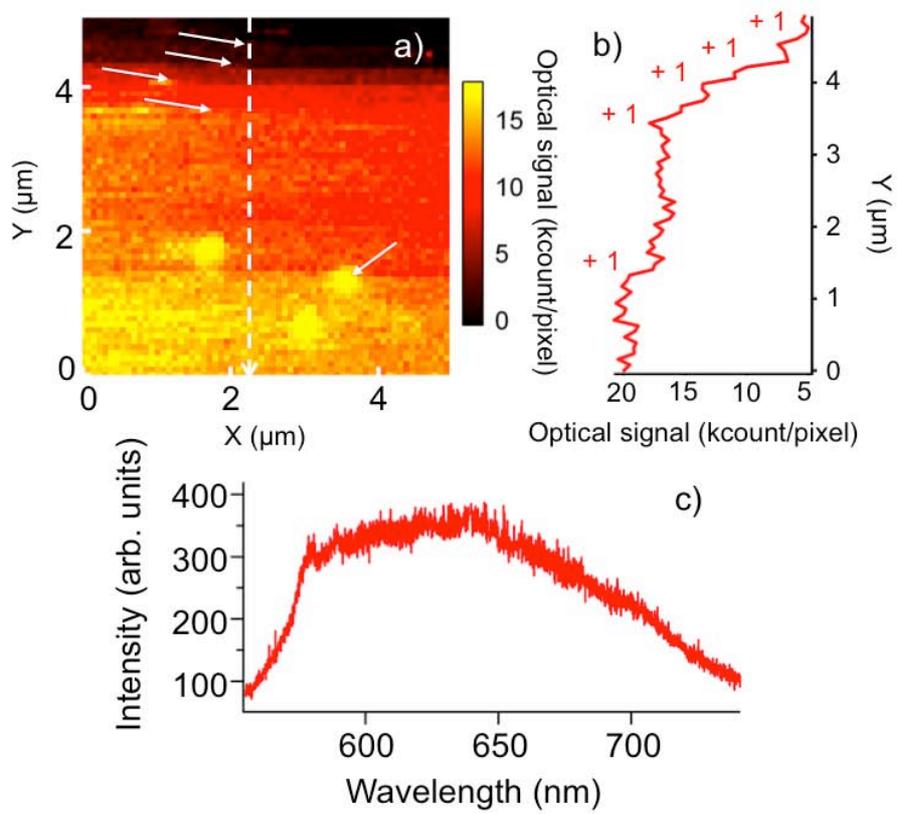

**Figure 1**

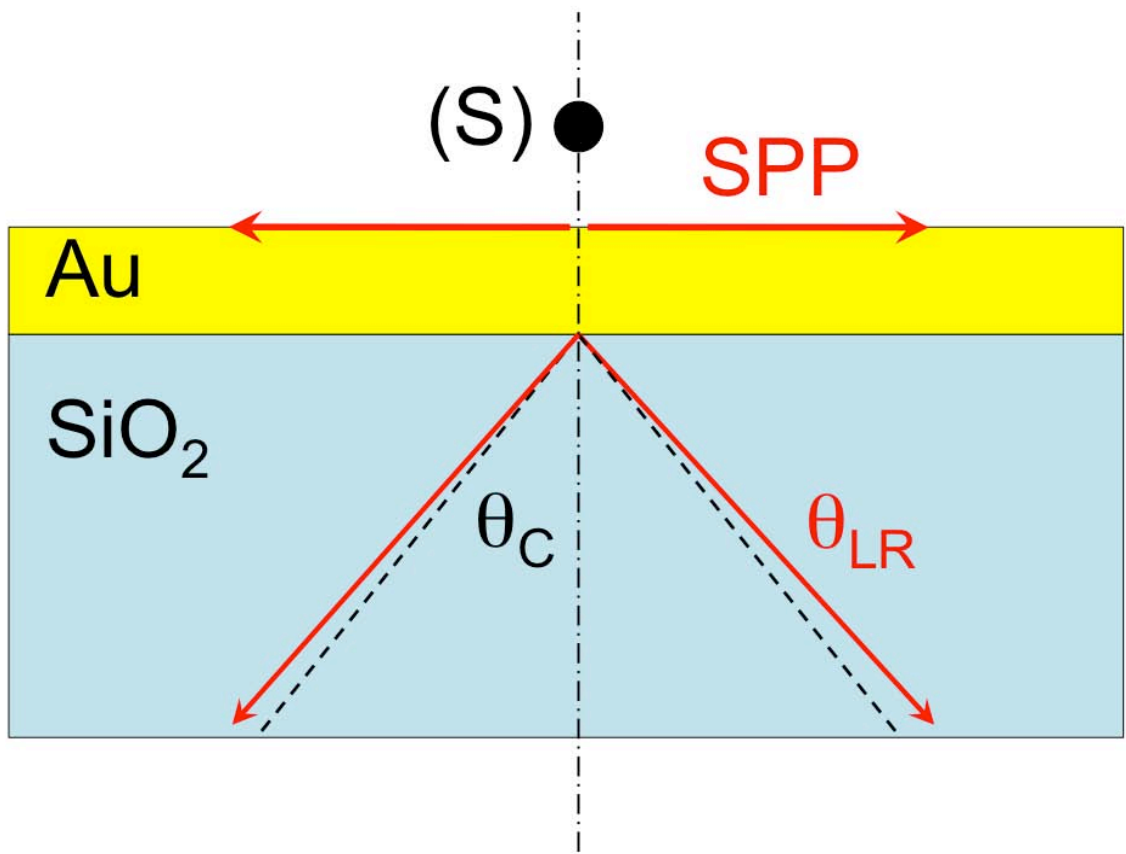

**Figure 2**

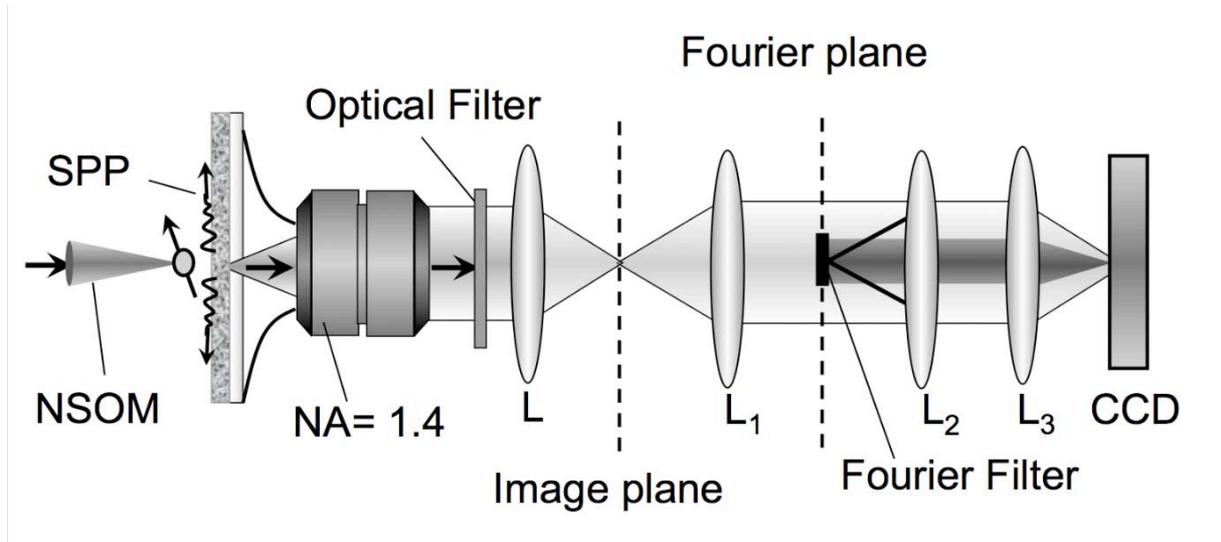

**Figure 3**

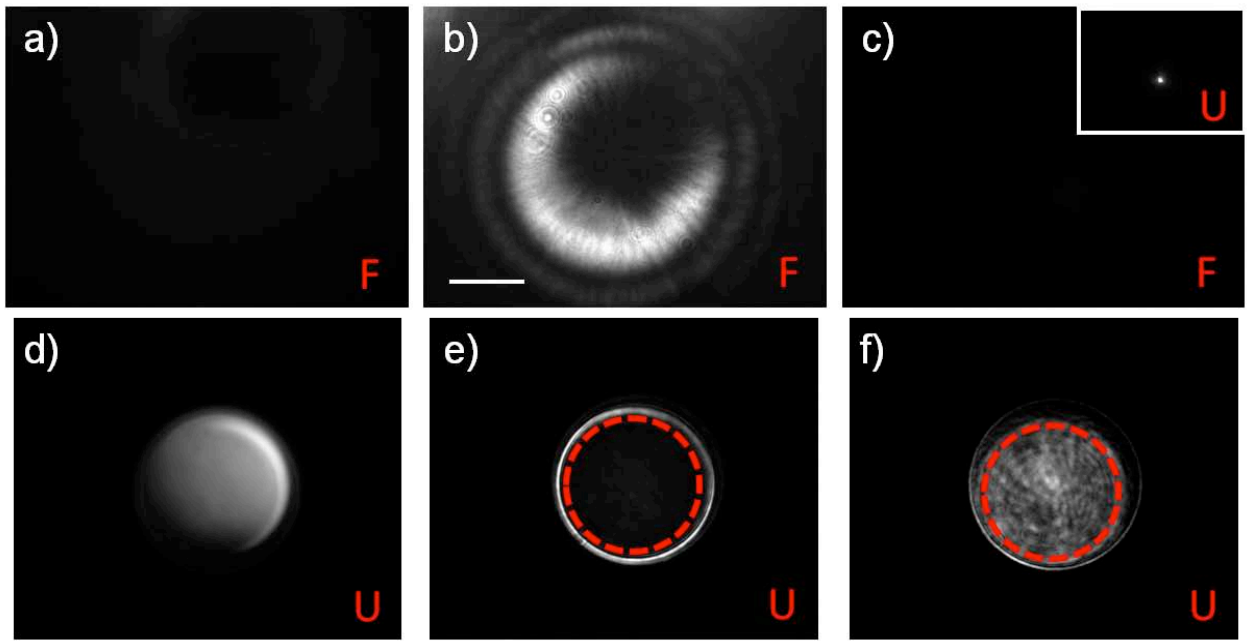

**Figure 4**

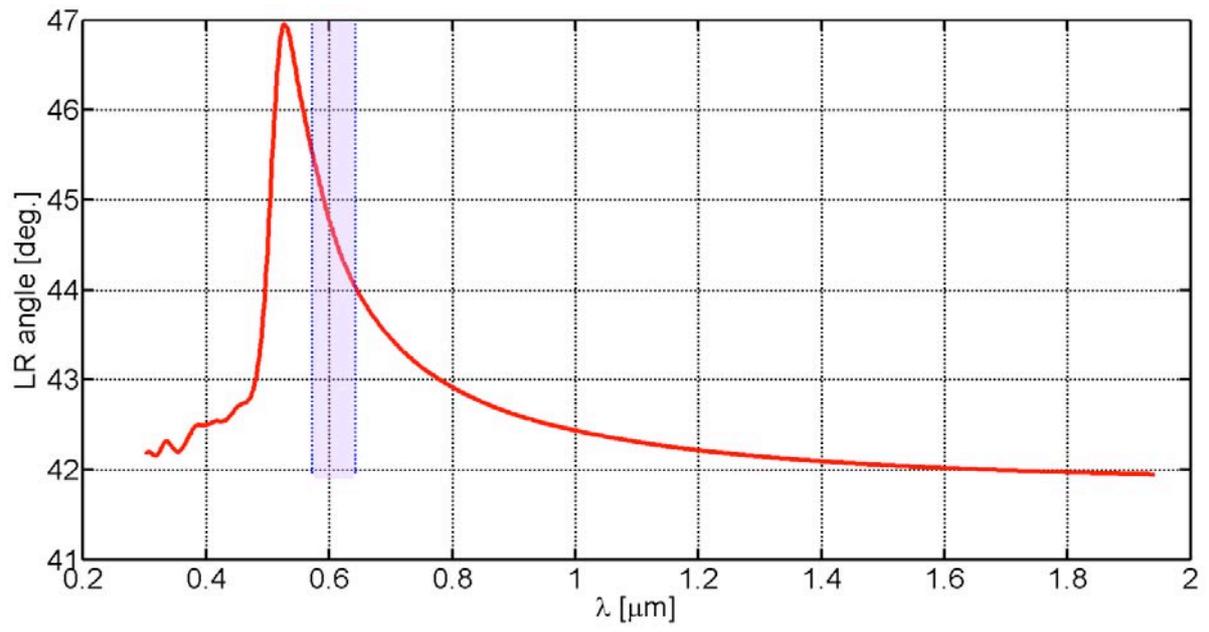

**Figure 5**